# End-to-End assessment of AR-assisted neurosurgery systems

Mahdi Bagheri, Farhad Piri, Hadi Digale, Saem Sattarzadeh, Mohammad Reza Mohammadi

**Abstract**

Augmented Reality (AR) has emerged as a significant advancement in surgical procedures, offering a solution to the challenges posed by traditional neuronavigation methods. These conventional techniques often necessitate surgeons to split their focus between the surgical site and a separate monitor that displays guiding images. Over the years, many systems have been developed to register and track the hologram at the targeted locations, each employed its own evaluation technique. On the other hand, hologram displacement measurement is not a straightforward task because of various factors such as occlusion, Vengence-Accomodation Conflict, and unstable holograms in space. In this study, we explore and classify different techniques for assessing an AR-assisted neurosurgery system and propose a new technique to systematize the assessment procedure. Moreover, we conduct a deeper investigation to assess surgeon error in the pre- and intra-operative phases of the surgery based on the respective feedback given. We found that although the system can undergo registration and tracking errors, physical feedback can significantly reduce the error caused by hologram displacement. However, the lack of visual feedback on the hologram does not have a significant effect on the user's 3D perception.

## Introduction

AR can be taken as the term referring to any case in which another reality is "augmented" to the environment by means of virtual computer graphic objects [1]. During the past years, AR opened a new area in different industries including game [2], medical applications [3], and architecture [4]. One of the most important means to implement the concept of AR is Head Mounted Display (HMD) which delegates a class of AR devices that can be worn on the head, expanding field vision producing imaginary screen appears to be several meters away [5] Microsoft HoloLens 2 (Microsoft, Redmond, WA) is an example of HMDs, incorporating a semi-transparent waveguided display alongside an on-board tracking system and multiple directly accessible sensors. This makes it a versatile device that can be used in many applications including healthcare and AR assisted surgeries[6]

Among AR applications, neurosurgery has attracted much attention due to the drawbacks of conventional methods [7]. With conventional methods, the surgeon has to look back and forth between the patient and a separate monitor, which can be mentally taxing as well as inaccurate. On the other hand, AR allows the surgeon to see guiding images and the patient's head simultaneously. However, bringing this technology into action has its own challenges to overcome. Vergence-Accommodation Conflict (VAC) impedes the eye from giving the best visual perception experience[8], lacking haptic feedback also prevents surgery simulation from being realistic [9], registering guiding images - which are usually represented by 3D-reconstructed CT and/or MRI objects - accurately on the targeted location is another challenge [10]. In addition, as HMDs are worn and moved in the space, it is important to track the device in the area to keep holograms in place where they were meant to be.

Registration is comprised of two major techniques: manual and automatic. Manual method can be done by moving the hologram to the proper position via hand interactions [11] and voice commands [12], or point-to-point fiducial-base registration [13]. However, automatic registration minimizes user intervention by utilizing data from HMD sensors [14].

Once registration is completed, the hologram appears in a fixed position relative to HMD's display unless a tracking system moves it in the opposite direction to keep it in place. Although Hololens 2 incorporates an advanced SLAM-based tracking system, it is prone to cause errors while walking or sudden head acceleration [15], causing the global coordinate system to drift away from its initial position and all the dependent holograms attached. To mitigate this, some studies suggest using external patterns to find Hololens pose. Kunz et al used several infrared markers in which by localizing them in space it will be possible to find Hololens pose [16].

When assessing registration and tracking, the primary challenge in measuring hologram displacement is to find a common reality knitting two worlds together accurately. Traditional measuring tools such as rulers are impractical because of the instability of holograms caused by inaccuracies in the tracking system, occlusion of holograms over every physical object, and blurriness caused by VAC. The purpose of this study is to standardize the assessment of AR-assisted neurosurgery systems by investigating current methods of registration and tracking assessments and propose a new method that can be easily and directly evaluated and conduct a deeper analysis of end-to-end system evaluation.

## Related works

Estimating the distance between the holographic and real world presents unique challenges due to the multitude of parameters involved and the difficulty in finding a device capable of estimating a reality in both worlds accurately. Most of the previous studies generally suggest 3 classes of methods for measuring the dislocation:

1. Using an external device [12, 13, 17-22],
2. Using a post-operative CT scan [23-26],
3. Harnessing the user's perspective [14, 27-29].

**External Device**

To adjust both realities, several studies used an external device to align either world into another or determine the same object's location in both (Fig. 1a). OptiTrack provides a Unity plug-in capable of being synced with Hololens, representing locations in a single coordinate system which makes it straightforward to measure the distance between where fiducial points are shown and where they should be shown [12]. Another study used the Polaris (Nothern Digital, Inc, Canada) tracking system as a gold standard to determine the ground truth points and ones that are distinguished by the user [17]. Meustee et al suggested a scenario to calibrate Hololens with a PST (PS-Tech, Amsterdam, Netherlands) tracking system to make it possible to visualize objects in 3D space on Hololens [18]. Gibby et al used NOVAPACS to measure target error in spine surgery [19]. Localization precision analysis was performed using the Micron tracker to record the position of the stylus tip for the user pointing at certain distinguishable fiducials on the 3D printed model. In the next stage fiducials on the virtual hologram were marked to compare and measure by point-wise Euclidean distance [20]. A stereo tracking system (NDI system) is used by Si et al to calibrate Hololens in physical space and then the user is asked to point at certain fiducials on a custom 3D-printed skull [21].

Andress et al adopted a different approach by creating a multimodality marker made part of metal to be able to be seen with a C-arm x-ray imaging system as well as Hololens. It has been evaluated through the measurement of four metal spheres and presented standard deviation as the yardstick for accuracy [22]. Another device is Aurora SCU which is used to determine the 6D catheter coordinates inside a patient's body in vascular surgeries [13].

Most of the methods that use external device require the user to point at specific fiducials with a stylus. The position of the stylus tip is then recorded and compared with the ground truth data which is obtained by the external device. The process can be done to determine hologram's fiducials positions [17] or both hologram's and model's [12, 21].

**Post-operative CT-scan**
Using CT-scan as the measurement tool, it is necessary to leave a distinguishable mark on the model which is visible in images (Fig. 1c). A drilling operation was performed by Liu et al under mixed reality guidance by Hololens, showing an arrow that converts to green when the drill is aligned with the guiding trajectory. Then the mark is left by a hole on the bone and with a post CT-scan imaging, direction and position deviations were measured [23]. Jiang et al adopted the same strategy in craniofacial surgery [24]. Another study used this method for EVD insertion in which the injected catheter tip is visible in post-operative CT-scan [25]. Spinal instrumentation under AR guide was performed by Muller et al and the translational and angular errors were reported in their paper [26]

**User's Perspective**
Some studies harnessed the user's own perspective to leave a mark either on a paper as a projection of the perspective or on the model itself (Fig. 1b). Using a hollow model and tracing borders with a pencil on a horizontal millimeter paper from predefined viewpoints is used to evaluate accuracy [27]. Another study evaluated the trajectory left on 3D printer models by showing a trail for the user [28]. A ruler is also used to measure the dislocation of the hologram and model [14]. Hologram displacement has also been measured by taking photos from different directions after registration. Then the dislocation that appeared in images was counted pixel-wise and converted to millimeter scale afterward [29].

**Comparison**
External device incorporation entails the provision of a device capable of either being synced with HMD or measuring a reality in both worlds. Although in some studies mean and standard deviation have been reported, in order to analyze end-to-end error, device accuracy and procedure precision should be considered too. In post-operative CT-scan, accuracy can be determined more easily just by knowing the accuracy of the equipment and image cuts because of the direct measurement approaches. However, to leave a mark visible in CT-scan images, there should be either an invasive operation [25] or a single-use model [23]. Among all, assessment based on the user's perspective is more prone to undergo errors in measurement due to hologram instability caused by tracking system movements and indefinite borders caused by VAC.

In this study we employ an external device to design a new method, that can be individually evaluated both in terms of device and procedure. This study not only evaluates target registration error for a custom system but also analyses the errors in different phases of the surgery:

1. While there is no interaction. Neither with the physical head nor with the hologram.
2. While interacting with holograms for planning.
3. While operating and using surgical tools under AR guidance.

*Table 1 – comparison of the methods*

| method | resolution | equipment | advantages | disadvantages |
|---|---|---|---|---|
| External device | • Between 0.08 mm and 0.5 mm in marker detection[30]<br>• 0.8mm in Arura[13]<br>• OptiTrack synced with Hololens N/A | Tracking device | • some can be synced with HMD.<br>• Finds markers in space. | • Tracking device provision.<br>• Device and procedure accuracy should be evaluated too. |
| Post-operative CT-scan | <=1mm | CT-scan equipment | • Digitized images can be simply evaluated.<br>• Both angular and translational deviations can be measured | • CT-scan device provision.<br>• Either makes the model single-use or requires an invasive surgery. |
| User's perspective | 1mm | - | | • Hologram instability and indefinite borders can harm precise data acquisition. |

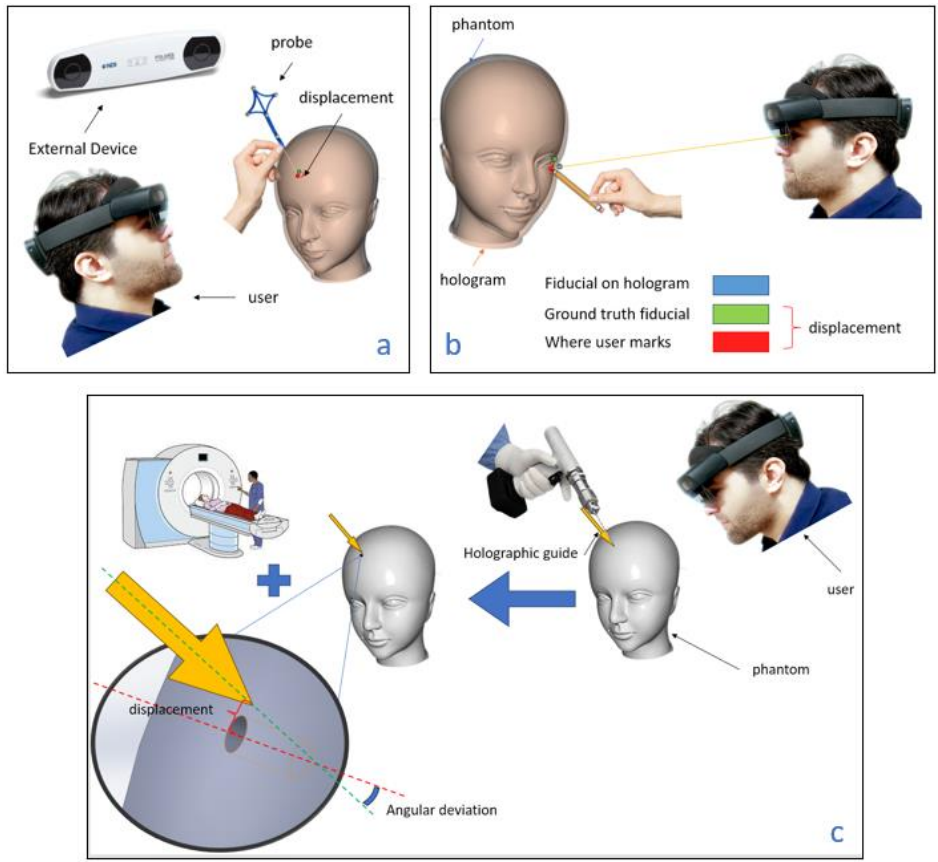

*Figure 1 a) schematic view of measurement by external device b) user's perspective c) post-operative CT scan*

## Material and Methods

Our error measurement method is based on the external device. Multiple Vicon cameras were calibrated in the Mowafaghian rehabilitation center gate analysis lab[1] (Fig. 2) where retroreflective markers could be tracked accurately. We used marker localization for two reasons. Firstly, to find markers installed on the probe to find the tip position, and secondly, to find markers installed on the phantom (Fig. 3) both in the lab coordinate system.

A Unity3D software was developed to acquire sensors' data and as user interface (UI) as well as to implement the tracking algorithm. A python software was developed as the server on a local laptop to do the computations for the registration algorithm. These two software were able to communicate with each other through a TCP connection. In the first step, the user registers the 3D-reconstructed CT-scan of the phantom and then marks certain fiducials blinking, one by one all over the head by placing the probe tip on them. The positions of the probe markers were recorded in the space and frames regarding the situation where user is pointing on fiducials were used to determine the accuracy by comparing them with ground truth data. The ground truth data was calculated accurately from data acquired from the phantom CT-scan and lab.

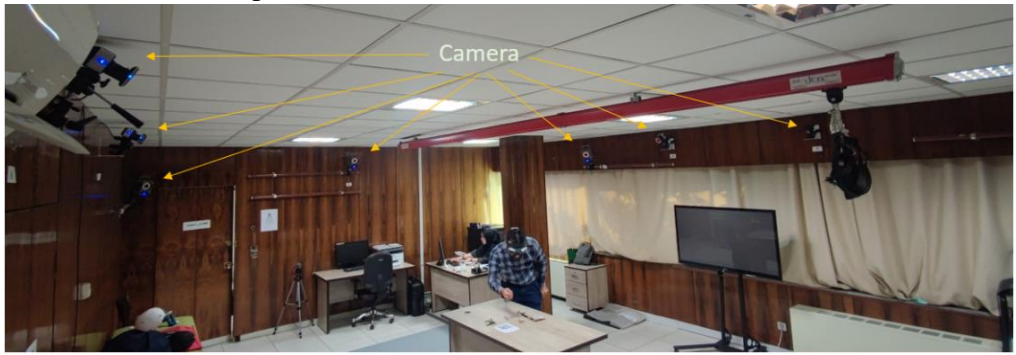

*Figure 2 wide view of the laboratory, showing cameras around the room*

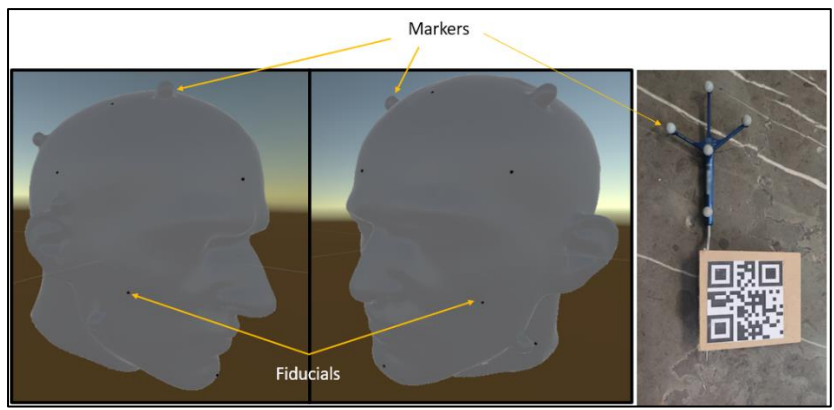

*Figure 3 two isometric views from CT scan in unity3D scene with fiducials and markers indicated alongside the probe and markers installed on it.*

---

[1] https://mowafaghian.ir/

Due to the importance of accurate estimation of probe tip coordinates, a CT-scan image with 1mm axial cut accuracy was captured from the probe and reconstructed afterward (Fig. 4a). The probe tip position was recorded as well (Fig. 4b) With 5 points in CT-scan coordinates and 5 acquired from the lab, we can find a point-to-point rigid transformation to convert the probe tip from the CT-scan coordinate to the lab coordinate in each frame. (Fig. 5)

To find the center of spheres visible in CT-scan of the probe (Fig. 4c), we used several vertices all over each, and estimated the center. We used least square optimization to fit an ideal sphere on selected points and calculated the center.

There were 9 retroreflective markers on the phantom all over the head (Fig. 4d). The center of each marker was determined using the method previously discussed. With 9 points on the reconstructed CT-scan of the phantom and 9 detected by the lab vision system, the rigid transformation was calculated with the same method as finding the transformation of the probe in the lab coordinate system. Then by applying calculated transformation on each fiducial on CT-scan, the ground truth location would be determined in the lab coordinate system. (Fig. 6)

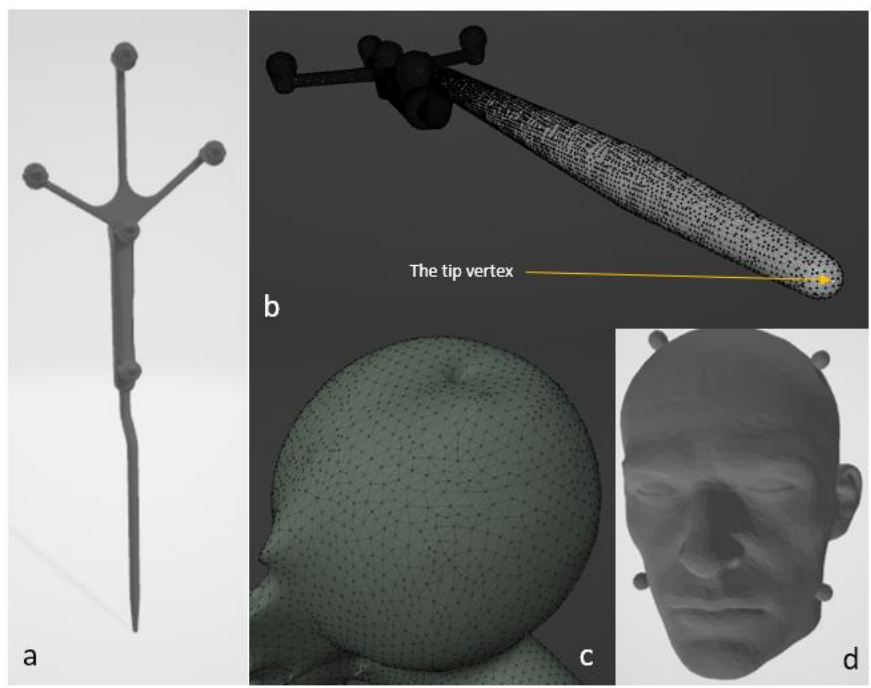

*Figure 4 a) reconstructed CT scan of the probe b) vertex regarding probe tip c) close view of a sphere CT scan d) reconstructed CT scan of the phantom*

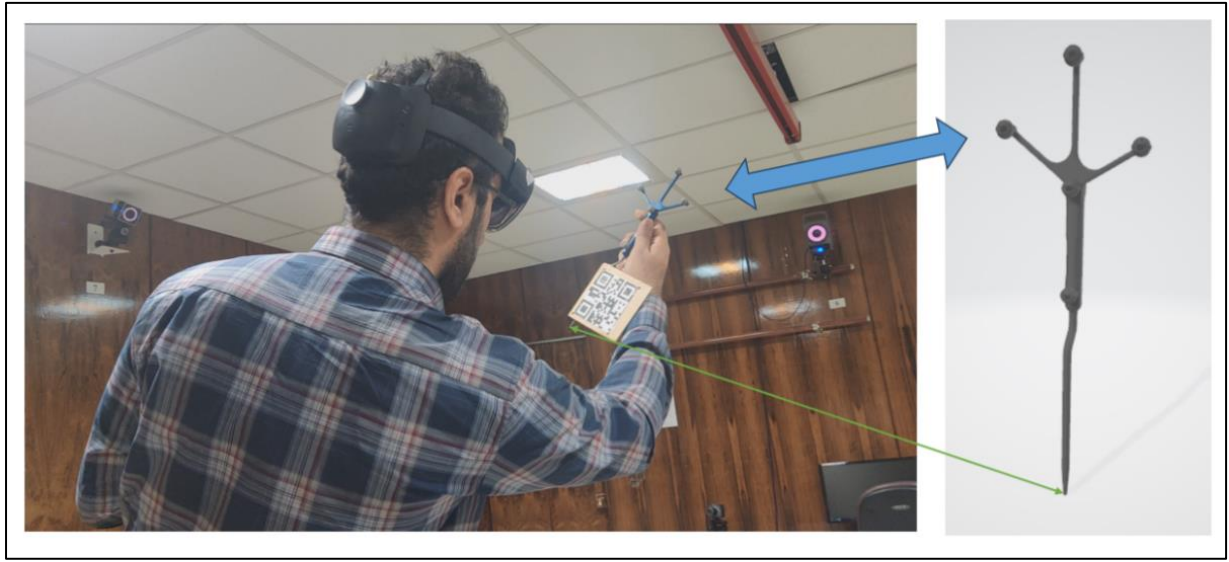

*Figure 5 Transformation between probe CT scan and lab*

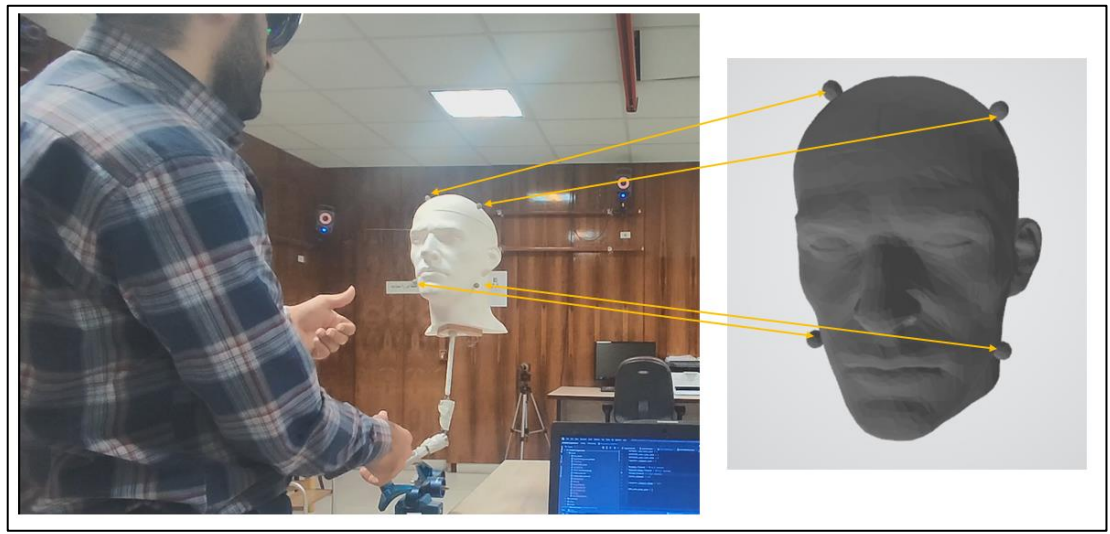

*Figure 6 transformation between phantom CT scan and lab*

**Tracking system**
Although Hololens 2 comes with a default tracking system, it is prone to drift error [15] which is significant in this context because it causes holograms to move away from where they have been registered. To mitigate this, we developed an additional algorithm that uses the main tracking system alongside a QR code which is fixed relative to the phantom. The Mixed Reality QR code tracking Unity plugin, provided by Microsoft[2] was used for QR code detection and localization. As soon as QR code localization and phantom registration are done, the relative transformation between them is recorded and considered as their fixed relative transformation. By using this method, even with Hololens global coordinate system drifting away, moved holograms would be relocated to the registered position as soon as the QR code is detected again.

---

[2] https://learn.microsoft.com/en-us/windows/mixed-reality/develop/advanced-concepts/qr-code-tracking-overview

**No feedback experiment**
Working with holograms does not provide the user with either haptic or visual feedback. Therefore, we designed this experiment as a baseline model to measure the surgeon's error based solely on observing the hologram. The process begins with performing registration. Then the phantom is moved away (without moving the QR code) which ends up with just the hologram floating in space. Next, the user marks fiducials one by one while estimating the depth based on his/her perception. (Fig. 7a)

**Holographic feedback**
In this experiment, the dislocation of the hologram was measured by providing visual feedback through the projection of the probe tip on the hologram's surface. During the pre-operative phase of the surgery, what the surgeon plans on, under the holographic guide is the hologram observed from the display. When interacting with the hologram (e.g. conical hole (Fig. 8a), plain cut (Fig. 8b), etc) user receives no feedback other than the interactions from the hologram. These feedbacks form his/her 3D perception of the inner segments of the brain. Therefore the hologram dislocation is individually important.

To simulate this error, holographic feedback was given to the user by calibrating a holographic arrow on the tip of the probe, tracked by another QR code fixed on the probe (Fig. 8c). The phantom was moved after registration and what was left in place was just the hologram (the same as no feedback) where it had been registered. In this experiment, as the probe tip approaches the surface, a proximity light appears and the user can see a colored pattern on the hologram surface which guides the user on how close the probe tip is to the hologram surface (Fig. 7b).

**Physical feedback experiment**
While operating under the AR guidance, the registered hologram has its own dislocation error. However, it does not necessarily equate to the surgeon's error while using surgical tools. If the dislocation is aligned with the user's viewpoint direction, because of the hologram occlusion, the fiducial position would not be determined by the user unless feedback is provided. In this case, the feedback is given when the probe tip hits the solid surface of the phantom. Thus, the location shown on the hologram is not necessarily where the operation will occur (Fig. 7c).

In this experiment, the hologram was registered on the phantom without being moved during the experiment. So, when the probe tip closes to the hologram, it will be stopped as soon as it hits the hard surface of the phantom and user marks the fiducials based on when the probe tip hits the surface.

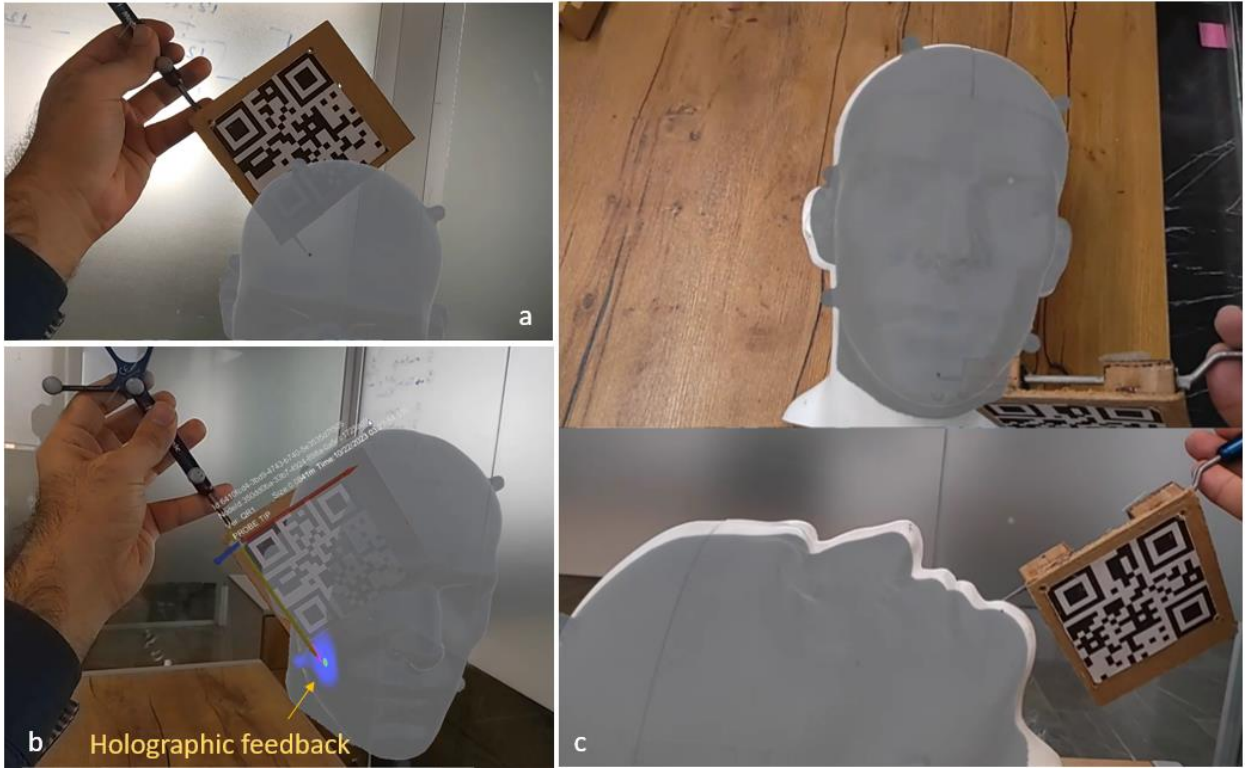

*Figure 7 different experiment types a) no feedback b) holographic feedback c) physical feedback when the prob tip is on the fiducial in front view and has a distance (closer to the correct point) in the side view.*

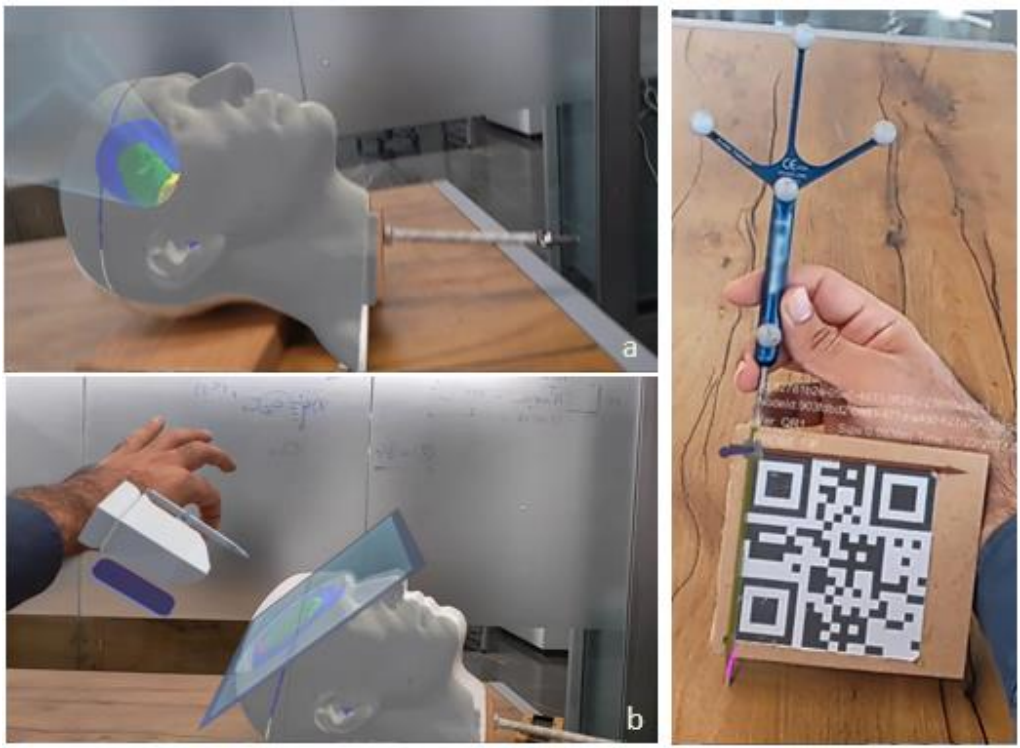

*Figure 8 a) displacement of hologram which gives a dislocated interpretation of inner segments for conical hole b) same for plain cut c) calibrated arrow (in magenta) on the probe tip*

## Results

In this section, we will first present the results related to the accuracy of sphere center estimation. Following this, we will discuss the transformation errors in identifying the probe tip, as well as fiducials on the phantom as ground truth. Subsequently, we will conduct Z-tests comparing holographic and physical feedback, as well as holographic and no-feedback experiments, to determine the significance of differences or similarities between distributions.

Totally 14 retroreflective spheres exist in experiments (5 attached to the probe and 9 on phantom) which in all cases center of them needed to be found in the lab as well as CT scan. To investigate the accuracy of sphere localization in the lab, we performed two experiments. Firstly, detection stability was investigated by putting spheres still on the ground and letting the system obtain multiple detections. Analyzing data revealed the standard deviation does not exceed 0.16mm for over 1400 frames. Secondly, as the probe is a solid object, distances between spheres should not change. However, while moving the probe in space we recorded spheres position and by comparing sphere-to-sphere distances, the maximum variance appeared to be 0.25mm.

On the other hand, to estimate the center of spheres in CT-scan (acquired with 1mm axial cut), least square estimation was used with several vertices all over the sphere (Fig. 4c). Standard deviation for this estimation was calculated to be 0.8mm at maximum.

With five points both from CT-scan (0.8mm error) and lab (0.25mm error), a rigid transformation was calculated for each frame (Fig. 5). The mean distance for corresponding points in experiments in which the probe tip was put on the hard surface was calculated 0.58 mm. The transformation was applied on the respective vertex for the tip to be found in the lab coordinate system. In such experiments, the variance in probe tip estimation did not exceed 1.09 mm (table 2 – physical feedback).

The same as the probe, we did calculations to find the transformation between phantom CT-scan and physical phantom under experiment (Fig. 6). Because spheres were scattered all over the head, the error regarding sphere estimation would be the same as the ground truth localization of fiducial error. After analyzing the data maximum standard deviation was no more than 1.51 mm (table 2 – gt error)

### Experiments results

Totally 3 sets of data were collected for each experiment type (physical, holographic, and no feedback) each indicating 16 fiducials. Results are shown in table 2.

*Table 2 results of the experiments*

| Experiment | trial | error(mm) | SD (mm) | Tip error (mm) | gt error (mm) |
|---|---|---|---|---|---|
| No feedback | 1 | 14.42 | 3.30 | 1.82 | 1.51 |
| | 2 | 11.81 | 2.31 | 1.85 | 1.50 |
| | 3 | 12.01 | 3.11 | 2.07 | 1.49 |
| | average | **12.75** | **2.94** | **1.92** | **1.50** |
| Holographic feedback | 1 | 10.01 | 2.12 | 1.78 | 1.51 |
| | 2 | 14.50 | 2.95 | 1.78 | 1.50 |
| | 3 | 11.47 | 3.71 | 1.52 | 1.51 |
| | average | **11.99** | **2.99** | **1.70** | **1.51** |
| Physical feedback | 1 | 6.86 | 2.19 | 0.85 | 0.91 |
| | 2 | 7.67 | 3.50 | 1.02 | 1.01 |
| | 3 | 6.40 | 3.28 | 1.09 | 1.01 |
| | average | **6.98** | **3.04** | **0.99** | **0.98** |

"error" indicates the mean distance of ground truth points and points indicated by the user. "SD" is the standard deviation of respective distances. "Tip error" is the standard deviation regarding probe tip estimation while the probe tip remained still to indicate a fiducial. Finally, "gt error" is the ground truth error regarding the transformation between CT-scan and lab which was discussed earlier.

**Tip error**

In the physical feedback experiment, the tip standard deviation is 0.99mm which in comparison to the other two experiments is relatively better. This is because, during physical feedback annotation, the tip of the probe gets fixed on a hard surface and remains still. However, in the other two, the user is annotating holograms on air, thus due to the hand trembling, the probe tip would also undergo some little movements.

**Comparisons**

Considering the null hypothesis as the physical and holographic error distributions are the same, by running a Z-test:

$$\sigma_{physical} = \frac{3.04}{\sqrt{48}} = 0.44, \sigma_{holographic} = \frac{2.99}{\sqrt{48}} = 0.43$$

$$Z = \frac{|\overline{X}_{physical} - \overline{X}_{holographic}|}{\sqrt{\sigma_{physical}^2 + \sigma_{holographic}^2}} = \frac{|6.98 - 11.99|}{0.61}$$

$$= 8.21 > 2 \rightarrow p \cong 0.0001$$

Thus, the null hypothesis can be rejected and distributions are significantly different. This means that in physical feedback, the user can mitigate the error in hologram dislocation by hitting the object surface with the surgical tool.

By performing Z-test on holographic feedback and no feedback experiment:

$$\sigma_{holographic} = 0.43, \sigma_{none} = 0.42$$

$$Z = = \frac{|6.98 - 11.99|}{0.61} = \frac{0.76}{0.6} = 1.27 \rightarrow p = 0.20$$

Hence, the null hypothesis cannot be rejected and the two distributions are the same. From this experiment, we can conclude that lack of depth feedback doesn't have a significant impact on user perception of the targeted points on hologram and user estimation is nearly identical when visual feedback is given.

## Discussion

**Methodology**

Several studies investigating methods for AR-guided neurosurgery for registration and/or tracking discussed their own methodology to assess their techniques. The lack of a unique, standard way to evaluate different assessment methods made us address this area. The current methods of using external devices discussed in the Related Works section, generally do not address the device and procedure accuracy and present the mean and standard deviation as the output. Nonetheless, understanding the accuracy of the assessment itself is crucial to comprehend the significance of the outputs.

Although using post-operative CT-scan benefits from direct measurement and its accuracy is the same as the accuracy of the device, despite the models being single-use or requiring invasive surgery, it can only measure what is left after surgery and is unable to address other errors happening in the middle of the operation (like holographic feedback we have discussed in this study).

Our proposed method benefits from being able to be individually evaluated. It is based on how accurately we can estimate markers in two coordinates (CT-scan and real world). Then the procedure involves finding point-to-point transformations between the center of spheres which can be easily assessed by the peer-wise distances. So, it is transparent at all levels, and ultimately, the significance of numbers can be determined.

Most of the previous works ( in all three types of methods) assessed the performance when physical feedback is given to the user. Although this is fine with the performance of the surgery, it doesn't show the registration error, but the mitigated situation where the user is receiving hints from other sources (phantom). To assess the registration, all the hints should be received from hologram, and then compared with ground truth. This was the motivation to introduce the holographic feedback experiment and it has been shown it is significantly different from physical feedback. However, it turned out that the user individually can have a good estimation of the depth of fiducials in spite of the occlusion of the hologram.

**Limitations**

The accuracy of this method depends on the accuracy of the external device and CT-scan in estimating the center of the spheres. This error naturally scales up when a procedure is applied to find other parameters (probe tip and fiducials). So, the more accurate the external device and CT-scan are, the more accurate the assessment results will be. Although our external device had promising accuracy in detecting the center of the spheres, CT-scan wasn't capable of estimating them as precisely and most of the uncertainties originated from there. Conventional CT-scan devices have below 0.234mm error [31] while ours was capable of 0.8mm at best. So, there is room for improvement by using a better device.

### Future works

Despite the improvements of using a better device to enhance the accuracy, this study only addressed registration errors in a limited time period while errors originating from the tracking system transpire over time. It is important to measure these errors because if we want to use such a system in surgery, the tracking system may undergo drift and holograms shift from where they had been originally registered. Thus, the accuracy in the early minutes wouldn't be the same as the end.

### Acknowledgment

This work is financially supported by Bina Teb Design & Development (Tehran, Iran).

### References

[1] P. Milgram and F. Kishino, "A taxonomy of mixed reality visual displays," *IEICE TRANSACTIONS on Information and Systems,* vol. 77, no. 12, pp. 1321-1329, 1994.

[2] J. Paavilainen, H. Korhonen, K. Alha, J. Stenros, E. Koskinen, and F. Mayra, "The Pokémon GO experience: A location-based augmented reality mobile game goes mainstream," in *Proceedings of the 2017 CHI conference on human factors in computing systems*, 2017, pp. 2493-2498.

[3] A. W. K. Yeung *et al.*, "Virtual and augmented reality applications in medicine: analysis of the scientific literature," *Journal of medical internet research,* vol. 23, no. 2, p. e25499, 2021.

[4] A. Cirulis and K. B. Brigmanis, "3D outdoor augmented reality for architecture and urban planning," *Procedia Computer Science,* vol. 25, pp. 71-79, 2013.

[5] T. Shibata, "Head mounted display," *Displays,* vol. 23, no. 1-2, pp. 57-64, 2002.

[6] S. Park, S. Bokijonov, and Y. Choi, "Review of microsoft hololens applications over the past five years," *Applied sciences,* vol. 11, no. 16, p. 7259, 2021.

[7] A. Meola, F. Cutolo, M. Carbone, F. Cagnazzo, M. Ferrari, and V. Ferrari, "Augmented reality in neurosurgery: a systematic review," *Neurosurgical review,* vol. 40, pp. 537-548, 2017.

[8] S. Condino, M. Carbone, R. Piazza, M. Ferrari, and V. Ferrari, "Perceptual limits of optical see-through visors for augmented reality guidance of manual tasks," *IEEE Transactions on Biomedical Engineering,* vol. 67, no. 2, pp. 411-419, 2019.

[9] C. Våpenstad, E. F. Hofstad, T. Langø, R. Mårvik, and M. K. Chmarra, "Perceiving haptic feedback in virtual reality simulators," *Surgical endoscopy,* vol. 27, pp. 2391-2397, 2013.

[10] L. Chen, T. W. Day, W. Tang, and N. W. John, "Recent developments and future challenges in medical mixed reality," in *2017 IEEE international symposium on mixed and augmented reality (ISMAR)*, 2017: IEEE, pp. 123-135.

[11] T. M. Gregory, J. Gregory, J. Sledge, R. Allard, and O. Mir, "Surgery guided by mixed reality: presentation of a proof of concept," *Acta orthopaedica,* vol. 89, no. 5, pp. 480-483, 2018.

[12] J. L. McJunkin *et al.*, "Development of a mixed reality platform for lateral skull base anatomy," *Otology & Neurotology: Official Publication of the American Otological Society, American Neurotology Society [and] European Academy of Otology and Neurotology,* vol. 39, no. 10, p. e1137, 2018.

[13] I. Kuhlemann, M. Kleemann, P. Jauer, A. Schweikard, and F. Ernst, "Towards X-ray free endovascular interventions–using HoloLens for on-line holographic visualisation," *Healthcare technology letters,* vol. 4, no. 5, pp. 184-187, 2017.

[14] A. Pepe *et al.*, "A marker-less registration approach for mixed reality–aided maxillofacial surgery: a pilot evaluation," *Journal of digital imaging,* vol. 32, pp. 1008-1018, 2019.

[15] R. Vassallo, A. Rankin, E. C. Chen, and T. M. Peters, "Hologram stability evaluation for Microsoft HoloLens," in *Medical Imaging 2017: Image Perception, Observer Performance, and Technology Assessment*, 2017, vol. 10136: SPIE, pp. 295-300.

[16] C. Kunz *et al.*, "Infrared marker tracking with the HoloLens for neurosurgical interventions," in *Current Directions in Biomedical Engineering*, 2020, vol. 6, no. 1: De Gruyter, p. 20200027.

[17] R. Moreta-Martinez, D. García-Mato, M. García-Sevilla, R. Pérez-Mañanes, J. Calvo-Haro, and J. Pascau, "Augmented reality in computer-assisted interventions based on patient-specific 3D printed reference," *Healthcare technology letters,* vol. 5, no. 5, pp. 162-166, 2018.

[18] J. W. Meulstee *et al.*, "Toward holographic-guided surgery," *Surgical*

[19] J. Gibby, S. Cvetko, R. Javan, R. Parr, and W. Gibby, "Use of augmented reality for image-guided spine procedures," *European Spine Journal,* vol. 29, pp. 1823-1832, 2020.

[20] T. Jiang, D. Yu, Y. Wang, T. Zan, S. Wang, and Q. Li, "HoloLens-based vascular localization system: precision evaluation study with a three-dimensional printed model," *Journal of medical Internet research,* vol. 22, no. 4, p. e16852, 2020.

[21] W. Si, X. Liao, Q. Wang, and P.-A. Heng, "Augmented reality-based personalized virtual operative anatomy for neurosurgical guidance and training," in *2018 IEEE Conference on Virtual Reality and 3D User Interfaces (VR)*, 2018: IEEE, pp. 683-684.

[22] S. Andress *et al.*, "On-the-fly augmented reality for orthopedic surgery using a multimodal fiducial," *Journal of Medical Imaging,* vol. 5, no. 2, pp. 021209-021209, 2018.

[23] H. Liu, E. Auvinet, J. Giles, and F. Rodriguez y Baena, "Augmented reality based navigation for computer assisted hip resurfacing: a proof of concept study," *Annals of biomedical engineering,* vol. 46, pp. 1595-1605, 2018.

[24] T. Jiang, M. Zhu, G. Chai, and Q. Li, "Precision of a novel craniofacial surgical navigation system based on augmented reality using an occlusal splint as a registration strategy," *Scientific reports,* vol. 9, no. 1, p. 501, 2019.

[25] Y. Li *et al.*, "A wearable mixed-reality holographic computer for guiding external ventricular drain insertion at the bedside," *Journal of neurosurgery,* vol. 131, no. 5, pp. 1599-1606, 2018.

[26] F. Müller, S. Roner, F. Liebmann, J. M. Spirig, P. Fürnstahl, and M. Farshad, "Augmented reality navigation for spinal pedicle screw instrumentation using intraoperative 3D imaging," *The Spine Journal,* vol. 20, no. 4, pp. 621-628, 2020.

[27] T. Frantz, B. Jansen, J. Duerinck, and J. Vandemeulebroucke, "Augmenting Microsoft's HoloLens with vuforia tracking for neuronavigation," *Healthcare technology letters,* vol. 5, no. 5, pp. 221-225, 2018.

[28] S. Condino *et al.*, "Wearable augmented reality platform for aiding complex 3D trajectory tracing," *Sensors,* vol. 20, no. 6, p. 1612, 2020.

[29] N. Q. Nguyen *et al.*, "An augmented reality system characterization of placement accuracy in neurosurgery," *Journal of Clinical Neuroscience,* vol. 72, pp. 392-396, 2020.

[30] A. Sorriento *et al.*, "Optical and electromagnetic tracking systems for biomedical applications: A critical review on potentialities and limitations," *IEEE reviews in biomedical engineering,* vol. 13, pp. 212-232, 2019.

[31] N. P. Dillon *et al.*, "Increasing safety of a robotic system for inner ear surgery using probabilistic error modeling near vital anatomy," in *Medical Imaging 2016: Image-Guided Procedures, Robotic Interventions, and Modeling*, 2016, vol. 9786: SPIE, pp. 438-452.